\documentclass[10pt,conference,letterpaper]{IEEEtran}
\usepackage{times,amsmath,epsfig}
\title{Time-aware Collaborative Filtering with the Piecewise Decay Function}
\author{%
{Pei Wu{\small $~^{1}$}, Chi-Ho Yeung{\small $~^{2}$}, Weiping Liu{\small $~^{1}$}, Cihang Jin{\small $~^{3}$}, Yi-Cheng Zhang{\small $~^{3}$} }%
\vspace{1.6mm}\\
\fontsize{10}{10}\selectfont\itshape
Faculty of Science, University of Fribourg\\
Fribourg, Switzerland\\
\fontsize{9}{9}\selectfont\ttfamily\upshape
$~^{1}$\{pei.wu, weiping.liu\}@unifr.ch\\
$~^{2}$phbill@ust.hk\\
$~^{3}$\{cihang.jin, zinppy\}@gmail.com%
}
\begin{document}
\maketitle
\begin{abstract}
In this paper, we determine the appropriate decay function for
item-based collaborative filtering (CF). Instead of intuitive
deduction, we introduce the Similarity-Signal-to-Noise-Ratio (SSNR)
to quantify the impacts of rated items on current recommendations.
By measuring the variation of SSNR over time, drift in user interest
is well visualized and quantified. Based on the trend changes of
SSNR, the piecewise decay function is thus devised and incorporated
to build our time-aware CF algorithm. Experiments show that the
proposed algorithm strongly outperforms the conventional item-based
CF algorithm and other time-aware algorithms with various decay
functions.

\end{abstract}

% NOTE keywords are not used for conference papers so do not populate them
\section{Introduction}

%Increasing efforts to incorporate temporal informations with
%collaborative filtering (CF) are observed in recent years, which is
%highly promising in raising accuracy and timeliness in
%recommendation.
On the rapidly changing Internet, user interest constantly changes
over time, which presents a unique challenge for practical
recommender systems. To cope with this, time-aware collaborative
filtering (CF) is proposed as a solution to provide timely
recommendations by exploiting the temporal information in
rating\footnote{Actually, various types of user behavior other than
rating exist, but in this paper, the term \emph{rate} and
\emph{rating} are loosely used to indicate all behaviors showing
user's preference, for example saving a bookmark, buying a product,
or voting on a movie.} data \cite{Ding:time}\cite{Koren:temporal},
such as the date the rating is generated. A straightforward and
low-complexity scheme is to incorporate, in the framework of
conventional CF, time-dependent weights in accounting the influence
of ratings with different ages. As rating influence generally
decreases with age, decay functions are introduced to weigh ratings
\cite{Ding:time}. Though benefits from time-awareness are observed
in several experiments \cite{Ding:time}\cite{Koren:temporal}, the
forms of decay function are still determined by intuitive deduction
because of limited understanding on temporal data. In this paper, we
identify and verify the appropriate form of decay function by
introducing a novel measure to quantify the influence of ratings,
which is the most crucial issue to understand and better utilize
temporal rating data. The major contributions are two-fold:

\begin{itemize}
\item[-]\emph{Revealing dynamics of individual rating behavior.}
A new index, the Similarity-Signal-to-Noise-Ratio (SSNR), is
introduced to quantify the impact of one's past ratings on his/her
current favorite. By measuring the SSNR on a real dataset with
precision in one second, we provide a clear picture of the rating
impact variation over time. Specifically, we observe three distinct
phases in the time variation of SSNR, namely a short-term decay, a
long-term decay and a plateau between them. By considering the
corresponding time scales, we suggest different mechanisms to
explain the short- and long-term decay respectively.
\item[-]\emph{The proposal of the piecewise decay function.}
By drawing analogy with signal combining, we apply the Maximal Ratio
Combining (MRC) method, and derive that the weights of ratings
should be proportional to their SSNR. Thus, the observed time
variation of SSNR provides a good reference for the decay function,
and the three distinct phases naturally lead to a piecewise power
decay function. Incorporating the decay function with CF leads to
our proposed time-aware recommendation algorithm.
\end{itemize}

To examine the recommendation quality,
% do you want to emphasize the word "experiment" here?
we test our proposed algorithm on the Delicious dataset, compared
with conventional item-based CF and other time-aware algorithms with
various decay functions. With \emph{hit-rate} as the evaluation
metric, remarkable improvements ranging from $11\%$ to $63\%$ are
observed.

\section{Related Work}

Time-aware CF algorithms are highly promising in raising
recommendation accuracy and timeliness. Among them, a kind of
recency-based algorithms, which employ the time window or the decay
function to emphasize recent ratings, are preferred for their
simplicity and adaptability to drifting of user interest. Reference
\cite{Ding:time} adopted the exponential function to produce
time-dependent weights for ratings. A more complex scheme was
proposed in \cite{Ding:recency}, which weighs ratings by their
prediction accuracy for recent ratings instead of by timestamps.
Furthermore, similar ideas were also used in \cite{Koren:temporal}
and \cite{Cao:enhance}, incorporated with more complex user interest
models. Besides the user interest drifting, studies in
\cite{Koren:temporal}\cite{Cao:enhance}\cite{xiang:graph} tried to
capture more temporal effects, but brought high computation burden
at the same time.

Besides in the time-aware CF, decay functions also play an important
role in many applications. An coarse-grained discrete function is
used in \cite{lee:implicity} to convert implicit ratings to
multi-value ratings. In \cite{Koychev:adaptation}, the linear
function is adopted to weigh examples for a content-based
recommender system. The most widely used decay function, exponential
function, is used in \cite{Klin:concept} for mining concept drifting
data.

\section{Temporal Dynamics in Ratings Data}
\subsection{Delicious Data}
\label{sec:del_data}

The rating data examined in subsequent analyses is collected from a
well known social bookmarking website, \emph{delicious.com}. Ratings
in the Delicious data are users' implicit feedbacks. When user
$\alpha$ saved bookmark (item) $i$ at time $t_{\alpha i}$, a
\emph{rating} occurred. To study the dynamics with a fine time
scale, the timestamp $t_{\alpha i}$ is recorded with precision in
one second.

After data collection, many bookmarks in the dataset are found to be
saved by only one user. These bookmarks have no contribution in
making recommendations, as they have no user overlap, and thus no
quantified relation with all other bookmarks. Hence we remove them
to avoid irrelevant computation. The attributes of the pre-processed
dataset are summarized in Table \ref{tab:datasets}. Note that the
sparsity of the Delicious data is very high and poses a big
challenge to recommendation algorithms \cite{xiang:graph}. As to
preserve its original features and make closer correspondence
between our study and the real world, we make no further
modification on the data.

\begin{table}[!t]
\centering
    \caption{Basic Information of the Delicious Data}     % NOTE!  caption goes _before_ the table contents !!
    \label{tab:datasets}

    \begin{small}
    \begin{tabular}{|l|l|l|l|l|}
    \hline
    {\bfseries Users} & {\bfseries Items} & {\bfseries
    Ratings} & {\bfseries Sparsity} & {\bfseries Period} \\ \hline
    14025 & 318415 & 1806951 & 0.9996 & Jan.2004 - Aug.2007\\\hline
    \end{tabular}
    \end{small}
\end{table}

\subsection{Temporal Dynamics of Rating Impact}

Intuitively impacts of ratings should decay with the lapse of time.
It implies that recent ratings are more relevant than old ratings to
identify one's current favorite. Many time-aware CF algorithms are
based on this deduction, but to the best of our knowledge, there is
no concrete evidence or clear picture about how the decay goes with
time. Limited knowledge on the decay process hinders the potential
improvements of the algorithms.

In this paper we address the problem of appropriate decay function
by empirical analyses on a real dataset. Firstly, a quantitative
measure is needed to quantify the rating impact on the current
recommendation. One candidate is the cosine similarity
\cite{Deshpande:item-based}, which is defined as
\begin{equation}
s_{ij} = \frac{\sum_{\alpha \in U} r_{\alpha i} r_{\alpha
j}}{\sqrt{\sum_{\alpha \in U} r_{\alpha i}^{2}} \sqrt{\sum_{\alpha
\in U} r_{\alpha j}^{2}}}, \label{eqn:cosine}
\end{equation}
where $s_{ij}$ is the similarity between items $i$ and $j$, and $U$
corresponds to the set of users. For the Delicious data, we adopt a
binary implicit rating and put $r_{\alpha i}=1$ if item $i$ is saved
by user $\alpha$ as a bookmark, and otherwise $r_{\alpha i} = 0$. We
further denote the current favorite item of user $\alpha$ as
$i_{\alpha}^*$, such that the similarity between his/her saved item
$i$ and $i_{\alpha}^*$ characterizes the impact of rating $r_{\alpha
i}$.

However, evaluating similarity is not sufficient to quantify rating
impact. Consider a simple case with two rated items $i$ and $k$ by
user $\alpha$. For item $i$, $s_{i i_{\alpha}^*} = 0.8$, but all
$s_{ij}$, where $j \neq k$ and $i_{\alpha}^*$, are larger than
$0.9$. For item $k$, $s_{k i_{\alpha}^*} = 0.2$, but all $s_{kj}$,
where $j \neq i$ and $i_{\alpha}^*$, are smaller than $0.1$. Though
$s_{i i_{\alpha}^*}>s_{k i_{\alpha}^*}$, apparently item $k$ is more
important than item $i$ to identify $i_{\alpha}^*$, the current
favorite of user $\alpha$. Therefore, the relative similarity with
the current item is more relevant than the sheer similarity.

To quantify the relative similarity with the favorite item, we draw
analogy with signal and noise in signal processing. Specifically,
$s_{i i_{\alpha}^*}$ can be considered as the useful signal carried
by the rated item $i$, and $s_{ij}$, where $j \neq i_{\alpha}^*$,
can be considered as noise. By drawing analogy with the standard
Signal-to-Noise-Ratio \cite{Boccuzzi:mrc}, we introduce the
Similarity-Signal-to-Noise-Ratio as
\begin{equation}
SSNR_{\alpha i} = \frac{s_{i i_{\alpha}^*}^2}{\sum_{j \neq
i_{\alpha}^*,j \neq i} s_{ij}^2}. \label{eqn:ssnr}
\end{equation}
High $SSNR_{\alpha i}$ implies item $i$ has a strong impact on
predicting the favorite item of user $\alpha$.

To investigate the decay of $SSNR_{\alpha i}$ with the age of rating
$r_{\alpha i}$, we conduct statistical analyses on the Delicious
data. For each user $\alpha$ in the dataset, we leave his/her latest
rating out, and consider the corresponding item as his/her current
favorite. The latest rated item and its timestamp are thus denoted
as $i_{\alpha}^*$ and $t_{\alpha}^*$. The SSNR of other rated items
are then evaluated by equation (\ref{eqn:ssnr}) with this favorite
item. For each rating of user $\alpha$ (except the latest one), a
pair of $(SSNR_{\alpha i}, age_{\alpha i})$, where $age_{\alpha i} =
t_{\alpha}^* - t_{\alpha i}$, is obtained and reveals the
relationship between rating impact and age. In this case, we will
compute $L-1$ pairs of $(SSNR_{\alpha i}, age_{\alpha i})$ for a
user with $L$ rated items, and the same computation is conducted for
all users.

\begin{figure}[h]
    \centerline{\psfig{figure=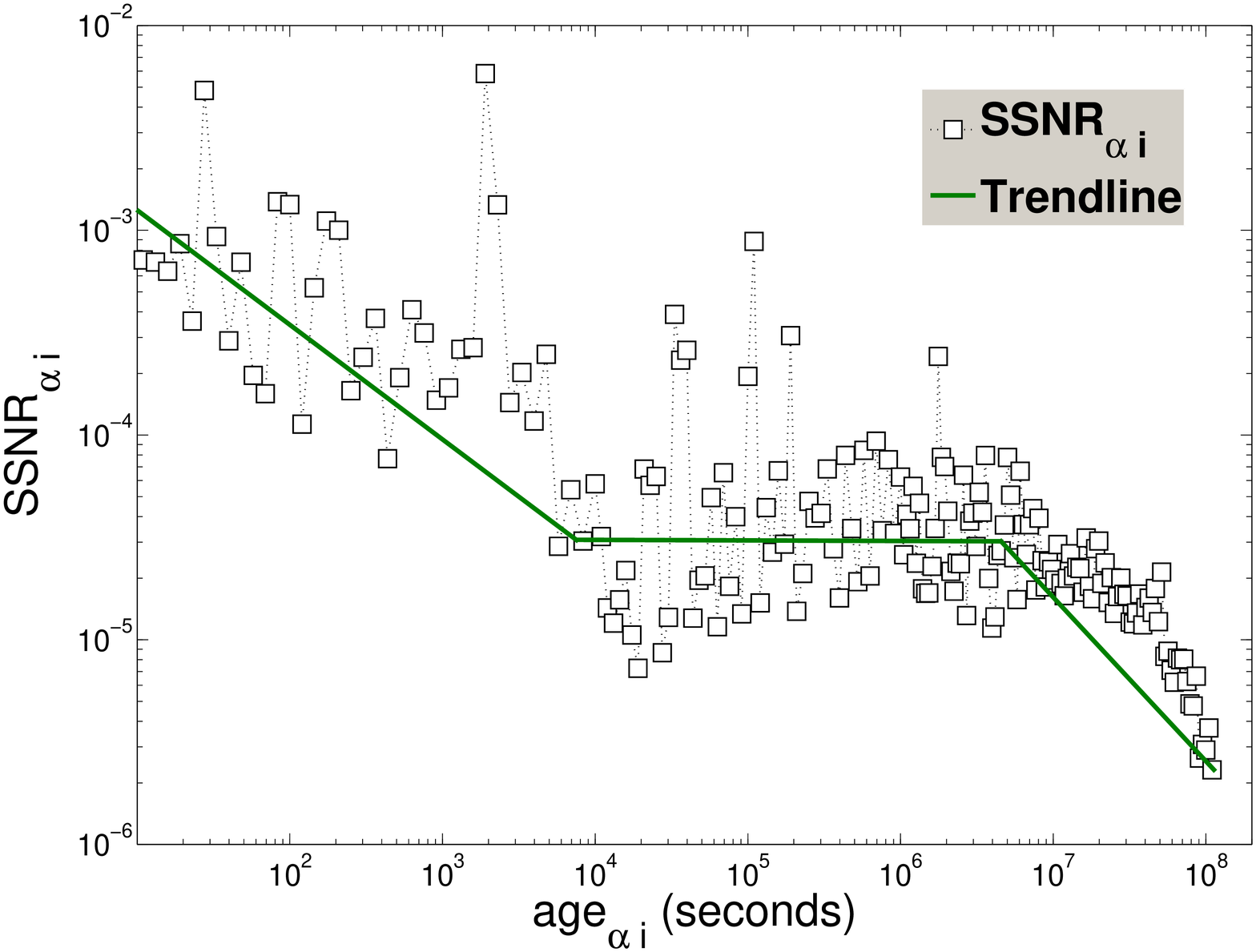,width=80mm} }
    \caption{Rating SSNR as a function of age. A trendline (the solid line) is added to outline the variation of SSNR.}
    \label{fig:ssnr}
\end{figure}

In Fig. \ref{fig:ssnr} we show $SSNR_{\alpha i}$ as a function of
$age_{\alpha i}$. We log-bin $age_{\alpha i}$ and average
$SSNR_{\alpha i}$ over each bin. Fig. \ref{fig:ssnr} is shown in
log-log scale to illustrate the behaviors of small $SSNR_{\alpha i}$
or with small $age_{\alpha i}$. A trendline is added to outline the
variation of SSNR.

%\begin{figure}[h]
%    \centering
%    \setlength\fboxsep{0pt}
%    \setlength\fboxrule{0.1pt}
%    \fbox{
%%    \centerline{\psfig{figure=ssnr_vs_age.eps,width=6mm} }}
%    \includegraphics[width=6.87cm]{ssnr_vs_age_new.eps}}
%    \caption{Rating SSNR as a function of ages. A trendline (dotted line) is added to outline the variation of SSNR.}
%    \label{fig:ssnr}
%\end{figure}

Despite fluctuation is relatively strong, the SSNR curve shows three
distinct phases. Specifically, \emph{short-} and \emph{long-term
decay} are respectively observed within $10^4$ seconds ($\approx 3$
hours) and beyond $10^6$ seconds ($\approx 10$ days), with a plateau
connecting the two phases. $SSNR_{\alpha i}$, defined as the impact
of rated item $i$ on current recommendation, is also a measurement
of how much interest of user $\alpha$ changes in time $age_{\alpha
i}$. Therefore, Fig. \ref{fig:ssnr} provides us a clear picture of
user interest drifting over time.

The corresponding time scales provide hints to explain the origin of
the decays. As neither individual nor global preference would have
great changes within $10^4$ seconds \cite{Koren:temporal}, we
attribute the \emph{short-term decay} to the switch of users'
focuses. As Fig. \ref{fig:ssnr} is shown, the values of $SSNR$
within $10^4$ are higher than the remains, which can indicate that
users' short-term focuses are more correlated. But the short-term
decay shows that users' focuses are drifting over time. By examining
a dataset with precision in one second, for the first time we
uncover the \emph{short-term decay} in rating data. In past studies,
these transient effects are regarded as interference to
recommendations and suggested to be filtered out
\cite{Koren:temporal}\cite{Cao:enhance}, but we will show in
following sections that they indeed have significant contributions
in improving recommendation accuracy.

The \emph{long-term decay} has been widely discussed in related
literature[1]--[4][7], which reflects the change of user's intrinsic
interest. Our quantitative results show that the change does not
occur in all time scales. It is only notable beyond about ten days,
but within this time window, user interest almost stays the same. We
suggest two main causes for the \emph{long-term decay}. The first
one is the attraction from new items, which constantly alters the
hotspots of the society. The second one is the change of users'
intrinsic characteristics, such as age, profession, social
relationship, etc. These changes happen slowly and less often, but
are very likely to affect users' preferences.

\section{Collaborative Filtering with the Piecewise Decay Function}

Exponential decay was proposed long ago to characterize the decay of
rating impact and is now widely used. It is based on human's
forgetting curve, which implies that the ratings decrease in impacts
as time goes on because people forget them. However, as discussed,
the temporal rating behaviors root in reasons in addition to
forgetting. In this section, we will derive an appropriate decay
function to incorporate with the item-based CF, which constitute our
proposed algorithm.

In general, recommendations by time-aware item-based CF are based on
% prediction function: is it a conventional term to use in this context?
\begin{equation}
%f_{\alpha j} = \sum_{i \in I_{\alpha}} w( age_{\alpha i} ) s_{ij},
% i have exchanged i and j to maintain consistency, see whether you like it or not
f_{\alpha j} = \sum_{i \in I_{\alpha}} w( age_{\alpha i} ) s_{ij},
\label{eqn:predict_time}
\end{equation}
where $f_{\alpha j}$ is the prediction score of item $j$ for user
$\alpha$, $I_{\alpha}$ denotes the set of all items rated by user
$\alpha$ and $w(t)$ is the decay function employed to weigh items
with different ages \cite{Ding:time}\cite{Deshpande:item-based}. For
user $\alpha$, all his/her unseen items are sorted by $f_{\alpha j}$
in the descending order, and the top-$N$ items are delivered as the
recommendation results.

In analogy with SSNR on similarity, we can also define SNR on the
final predicted probabilities as
%$SSNR_{f} = f_{\alpha i_{\alpha}^*} / \sum_{k \neq i_{\alpha}^*} s_{ik}$,
% i have used fSNR (f-signal-to-noise) ratio, and added the alpha dependence on fSNR
%see whether you like it or not
$fSNR_{\alpha} = f_{\alpha i_{\alpha}^*}^2 / \sum_{k \neq
i_{\alpha}^*} f_{\alpha k}^2$, where $i_{\alpha}^*$ is again the
current favorite of user $\alpha$. Obviously, good recommendation
results should give item
%$\alpha$, see whether you would like to mean $i_{\alpha}^*$  here
$i_{\alpha}^*$ a high rank and hence a large $fSNR_{\alpha}$.
%Then the
%problem is, how to arrange the weight for each similarity to
%maximize the SSNR of their weighted sum.
We then have to assign appropriate weights, i.e. decay function, to
maximize the outcoming $fSNR$. By drawing analogy with signal
combing problem, where MRC is employed for the same
purpose\footnote{MRC is the optimal combiner for independent
Additive-White-Gaussian-Noise channels. For other channel types, MRC
is also widely adopted , because its basic idea of boosting the
strong signal components and attenuating the weak components will
surely improve performance when compared with Equal Gain Combing.},
we obtain $w( age_{\alpha i} ) = SSNR_{\alpha i}$. Based on the
variation of $SSNR_{\alpha i}$ with $age_{\alpha i}$ in Fig.
\ref{fig:ssnr}, we take the trendline of the curve to derive $w(
age_{\alpha i} )$. Mathematically, the decay function reads
\begin{equation}
w( t ) = \left\{
\begin{array}{cl}
\left( \frac{t}{T_s} \right)^{-K_s}, & 0 \leq t < T_s, \\
1, & T_s \leq t < T_l, \\
\left( \frac{t}{T_l} \right)^{-K_l}, & T_l \leq t.
\end{array}
\right. \label{eqn:decay_del}
\end{equation}
Substituting (\ref{eqn:decay_del}) into (\ref{eqn:predict_time}), we
propose a time-aware CF algorithm with the piecewise decay function.
In equation (\ref{eqn:decay_del}), four free parameters, $T_s$,
$T_l$, $K_s$ and $K_l$, are introduced for fine tuning to achieve
the optimal algorithmic performance. $T_s$ and $T_l$ are
respectively the time thresholds of short- and long-term decay, and
$K_s$ and $K_l$ are the control parameters for the corresponding
decay rate.

\begin{table*}[!ht]
%\centering
    \begin{minipage}[!ht]{0.7\linewidth}
    \centering
    \caption{Algorithms' Hit-Rate and Improvements}     % NOTE!  caption goes _before_ the table contents !!
    \label{tab:result}
    \begin{small}
    \begin{tabular}{|c|cc|cc|cc|}
    \hline
    {\bfseries Algorithms} & {$H@10$} & {Improvement} & {$H@20$} & {Improvement} & {$H@50$} & {Improvement}\\
     & {\bfseries ($\times 10^{-3}$)} & & {($\times 10^{-4}$)} & & {($\times 10^{-4}$)} & \\
    \hline\hline
    IBCF & $1.03$ & $-$ & $6.84$ & $-$ & $4.23$ & $-$\\
    \hline
    WIN & $1.29$ & $\uparrow 25.2\%$ & $8.23$ & $\uparrow 20.3\%$ & $4.68$ & $\uparrow 10.6\%$\\
    \hline
    LOG & $1.44$ & $\uparrow 39.8\%$ & $8.71$ & $\uparrow 27.3\%$ & $4.59$ & $\uparrow 8.5\%$\\
    \hline
    EXP & $1.51$ & $\uparrow 46.6\%$ & $9.00$ & $\uparrow 31.6\%$ & $4.66$ & $\uparrow 10.2\%$\\
    \hline
    OUTRADAY & $1.54$ & $\uparrow 49.5\%$ & $9.74$ & $\uparrow 42.4\%$ & $5.48$ & $\uparrow 29.6\%$\\
    \hline
    {\bfseries Proposed} & $\mathbf{1.68}$ & $\uparrow \mathbf{63.1\%}$ & $\mathbf{10.6}$ & $\uparrow \mathbf{55.0\%}$ & $\mathbf{5.76}$ & $\uparrow \mathbf{36.2\%}$\\
    \hline
    \end{tabular}
    \end{small}
    \end{minipage}%
    \begin{minipage}[!ht]{0.3\linewidth}
    \centering
    \caption{Optimal Parameters}     % NOTE!  caption goes _before_ the table contents !!
    \label{tab:parameters}
    \begin{small}
    \begin{tabular}{|c|c|}
    \hline
    $T_w$ & $10^7$ \\\hline
    $T_g$ & $30000$ \\\hline
    $T_e$ & $50000$ \\\hline
    $K_o$ & $0.9$ \\\hline
    $T_s$ & $50000$ \\\hline
    $T_l$ & $10^6$ \\\hline
    $K_s$ & $0.6$ \\\hline
    $K_l$ & $0.3$ \\\hline
    \end{tabular}
    \end{small}
    \end{minipage}
\end{table*}

\section{Experiments}

\subsection{Experiment Design}

We evaluate our algorithm on the Delicious data described in section
\ref{sec:del_data}. As the task of recommender systems is to
identify one's \emph{current} favorite, we adopt the so-called
leave-the-latest-out method for cross validation, rather than the
traditional K-fold or leave-one-out method. The latest rating of
each user, say user $\alpha$, is left out as a probe when making
recommendations for him/her, and all the other ratings serve as the
training data. The timestamp $t_{\alpha}^*$ of the latest rating is
regarded as the time when the recommendation is made, and is used to
calculate ratings' ages. The latest rated item $i_{\alpha}^*$ is the
test item for evaluating the recommendation accuracy.

In this paper, \emph{hit-rate} is employed as the evaluation metric
for recommendation accuracy \cite{Deshpande:item-based}. The
\emph{hit-rate} for a specific search depth $N$ is defined as
\begin{equation}
H@N = \frac{1}{|U|} \sum_{\alpha \in U}\frac{h(i_{\alpha}^*, N)}{N},
\label{eqn:precision}
\end{equation}
where $|U|$ is the number of users, $h(i_{\alpha}^*, N)=1$ if item
$i_{\alpha}^*$ is top-$N$ sorted by prediction scores, and
$h(i_{\alpha}^*, N) = 0$ otherwise.

\subsection{Decay Functions}
The proposed algorithm is simulated with the parameters $K_s \in
[0.1, 1]$, $K_l \in [0.1, 1]$, $T_s \in [100, 10^5] $\footnote{In
this paper, all values of time is in unit of one second.} and $T_l
\in [5 \times 10^5, 5\times 10^7]$. To demonstrate the benefits from
temporal information, we include the conventional Item-Base CF
(IBCF) in our experiment. Besides, we also evaluate time-aware CF
with the following decay functions for comparison.

\begin{itemize}
\item \emph{WINdow function} (WIN):
\begin{equation}
w_{w}(t) = \left\{ \begin{array}{ll}1, & \textrm{$ t \leq
T_{w}$}\\0, & \textrm{$ t > T_{w}$}\end{array}\right. ,
\label{eqn:timewindow}
\end{equation}
where the free parameter $T_w \in [100, 10^8]$.
\item \emph{LOGistic function} (LOG) \cite{Ding:time}:
\begin{equation}
w_{g}(t) = \frac{1}{1 + \exp(\frac{t}{T_{g}}- b)},
\label{eqn:logistic}
\end{equation}
where the free parameter $T_g \in [1, 10^8]$, and $b$ is set to $5$
in our experiment.
\item \emph{EXPonential function} (EXP) \cite{Ding:time}:
\begin{equation}
w_{e}(t) = \exp(-\frac{t}{T_{e}}), \label{eqn:exponential}
\end{equation}
where the free parameter $T_e \in [1, 10^8]$.
\item \emph{OUTRADAY decay function} (OUTRADAY):
\begin{equation}
w_{o}(t) = \left\{ \begin{array}{cl}1, & \textrm{$ t < 86400$}\\
\left( \frac{t}{86400} \right)^{-K_o}, & \textrm{$ t \geq
86400$}\end{array}\right. , \label{eqn:timewindow}
\end{equation}
where the free parameter $K_o \in [0.1, 2]$. The outraday decay
function is similar to the proposed decay function, except that it
ignores the short-term decay which happens within one day.
\end{itemize}

\subsection{Results}
In our experiment, for each algorithm we calculate many groups of
$H@10$, $H@20$ and $H@50$ by tuning the free parameters in the given
regions. Then, the results with optimal $H@10$ are selected and
shown in Table \ref{tab:result}, and at the same time the optimal
parameters are given in Table \ref{tab:parameters}. We also list the
improvements of each algorithm when compared with the IBCF
algorithm. Apparently, all time-aware algorithms strongly outperform
the IBCF algorithm. Among them, the proposed algorithm achieves the
best performance, and the improvements are significant. Remarkably,
the difference in improvement between the proposed and OUTRADAY
algorithm shows the great importance of short-term decay, as the two
algorithms are identical except that the latter ignores the dynamics
within one day. This again confirms the importance of the present
study, as most past studies overlook the benefits of examining
dynamics shorter than one day
\cite{Koren:temporal}\cite{xiang:graph}.

\section{Conclusions and Discussion}

In this paper, we quantified user interest drifting by a novel
quantitative measure by analogy with Signal-to-Noise-Ratio, and
applied the findings on our time-aware CF algorithm by a carefully
designed decay function. We uncovered and utilized the short-term
decay shorter than one day, which is overlooked in the past studies.
Experiments show our great algorithmic improvement compared with the
present state-of-the-art.

It is worth noting that, users' activities in the Internet are more
bursty than what we observed from the Delicious data, where rich
short-term information is ready to be utilized to improve
recommendation. As a further example, we construct a semi-artificial
dataset from the Delicious data, which emphasizes the bursty
behaviors. Experiments on this dataset demonstrate that an
improvement of $110\%$ is achieved by the proposed algorithm when
compared with the IBCF algorithm. The results will be presented in
details in an extended paper.

%inferred the appropriate form for item-based CF, and built a
%time-aware recommender system based on the piecewise power decay
%function. With the picture of SSNR's variance versus age, we
%revealed three distinct phases in the temporal dynamics of
%individual's rating behavior. Among them, the short-term decay and
%the stable stage of users interest have never been uncovered before
%due to the use of coarse time resolution and the lack of suitable
%analysis method.

\section*{Acknowledgment}
This work is supported by the Liquid Publications project (EU
FET-Open Grants 213360).

% the following vfill coursely balances the columns on the last page
\vfill

%\pagebreak

\bibliographystyle{IEEEtran}

\bibliography{IEEEabrv,IEEEexample}

\begin{thebibliography}{10}
\providecommand{\url}[1]{#1}
\csname url@rmstyle\endcsname
\providecommand{\newblock}{\relax}
\providecommand{\bibinfo}[2]{#2}
\providecommand\BIBentrySTDinterwordspacing{\spaceskip=0pt\relax}
\providecommand\BIBentryALTinterwordstretchfactor{4}
\providecommand\BIBentryALTinterwordspacing{\spaceskip=\fontdimen2\font plus
\BIBentryALTinterwordstretchfactor\fontdimen3\font minus
  \fontdimen4\font\relax}
\providecommand\BIBforeignlanguage[2]{{%
\expandafter\ifx\csname l@#1\endcsname\relax
\typeout{** WARNING: IEEEtran.bst: No hyphenation pattern has been}%
\typeout{** loaded for the language `#1'. Using the pattern for}%
\typeout{** the default language instead.}%
\else
\language=\csname l@#1\endcsname
\fi
#2}}

\bibitem{Ding:time}
Y.~Ding and X.~Li, ``Time weight collaborative filtering,'' in \emph{Proc. ACM
  CIKM '05}, Bremen, Germany, Oct. 2005, pp. 485--492.

\bibitem{Koren:temporal}
Y.~Koren, ``Collaborative filtering with temporal dynamics,'' in \emph{Proc.
  ACM KDD '09}, Paris, France, June 2009, pp. 447--456.

\bibitem{Ding:recency}
Y.~Ding, X.~Li, and M.~E. Orlowska, ``Recency-based collaborative filtering,''
  in \emph{Proc. ADC '06}, Hobart, Australia, Jan. 2006, pp. 99--107.

\bibitem{Cao:enhance}
H.~H. Cao, E.~H. Chen, J.~Yang, and H.~Xiong, ``Enhancing recommender systems
  under volatile userinterest drifts,'' in \emph{Proc. ACM CIKM '09}, Hong
  Kong, China, Nov. 2009, pp. 1257--1266.

\bibitem{xiang:graph}
L.~Xiang, Q.~Yuan, S.~W. Zhao, L.~Chen, X.~T. Zhang, and J.~M. Sun, ``Temporal
  recommendation on graphs via long- and short-term preference fusion,''
  \emph{in Proc. ACM KDD '10}, July 2010, to be published.

\bibitem{lee:implicity}
T.~Q. Lee, Y.~Park, and Y.~T. Park, ``A time-based approach to effective
  recommender systems using implicit feedback,'' \emph{Expert Systems with
  Applications: An International Journal}, vol.~34, pp. 3055--3062, May 2008.

\bibitem{Koychev:adaptation}
I.~Koychev and I.~Schwab, ``Adaptation to drifting user's interests,'' in
  \emph{Proc. ECML Workshop '00}, Barcelona, Spain, May 2000, pp. 39--45.

\bibitem{Klin:concept}
R.~Klinkenberg and S.~R¨¹ping, ``Concept drift and the importance of
  examples,'' in \emph{Text Mining¨CTheoretical Aspects and Applications},
  J.~Franke, G.~Nakhaeizadeh, and I.~Renz, Eds.\hskip 1em plus 0.5em minus
  0.4em\relax Berlin, Germany: Springer, 2003, pp. 55--77.

\bibitem{Deshpande:item-based}
M.~Deshpande and G.~Karypis, ``Item-based top-n recommendation algorithms,''
  \emph{ACM Trans. Inf. Syst.}, vol.~22, pp. 143--177, Jan. 2004.

\bibitem{Boccuzzi:mrc}
J.~Boccuzzi, \emph{Signal Processing for Wireless Communications}.\hskip 1em
  plus 0.5em minus 0.4em\relax New York, NY: McGraw-Hill Professional, 2008.

\end{thebibliography}

\end{document}